\documentstyle[11pt,epsf]{article}

\newcommand{\beq}{\begin{equation}}
\newcommand{\eeq}{\end{equation}}
\newcommand{\beqa}{\begin{eqnarray}}
\newcommand{\eeqa}{\end{eqnarray}}

\newcommand{\PLB}[1]{{\it Phys. Lett.}\ {\bf B{#1}}}
\newcommand{\PRD}[1]{{\it Phys. Rev.}\ {\bf D{#1}}}

\newcommand{\tcii}{${\rm TC}^2$ }

\def\st{\sin\theta}
\def\ct{\cos\theta}
\def\sp{\sin\phi}
\def\cp{\cos\phi}

\begin{document}

\begin{titlepage}
\def\thepage {}        

\title{Precision Electroweak Constraints on Top-Color Assisted Technicolor}

\author{
R.S. Chivukula\thanks{e-mail addresses: sekhar@bu.edu, terning@calvin.bu.edu},
and J. Terning$^*$\\
Department of Physics, Boston University, \\
590 Commonwealth Ave., Boston MA  02215}

\date{June 4 1996}

\maketitle

\bigskip
\begin{picture}(0,0)(0,0)
\put(295,250){BUHEP-96-12}
\end{picture}
\vspace{24pt}

\begin{abstract}
Using precision electroweak data, we put limits on ``natural''
top-color assisted technicolor models. Generically the
new $U(1)$ gauge bosons in these models must have masses
larger than roughly 2 TeV, although in certain (seemingly unrealistic) models
the bound can be much lower.
\pagestyle{empty}
\end{abstract}
\end{titlepage}

\section{Introduction}
\label{sec:intro}
Recently Hill \cite{tc2} introduced top-color assisted technicolor
(${\rm TC}^2$) in order to explain electroweak symmetry breaking via a
dynamical mechanism and to allow for a large top quark mass.  In this
model a top-condensate is driven by the combination of a strong
isospin-symmetric top-color interaction and an additional (possibly weak,
but probably strong) isospin-breaking $U(1)$ interaction.  He argued that the
extreme fine-tuning that was required in pure top-condensate models
can be done away with if the scales of the critical top-color and
$U(1)$ interactions are brought down to a TeV.  Given a top quark mass
of around 175 GeV, such top-color interactions would produce masses
for the $W$ and $Z$ that are far too small, hence there must be
further strong interactions (technicolor) that are primarily
responsible for breaking electroweak symmetry.

The authors of ref. \cite{tc2isospin} argued that the isospin breaking
$U(1)$ gauge interactions, which are necessary in order to split the
top and bottom quark masses, are likely to have isospin-violating
couplings to technifermions.  In this case they produce a significant
shift in the $W$ and $Z$ masses (i.e.~contribute to $\Delta \rho_* =
\alpha T$).  In order to satisfy experimental constraints on $\Delta
\rho_*$, they found that either the effective top quark coupling or the
top-color coupling must be tuned to 1\%.

Subsequently, Lane and Eichten \cite{naturaltc2} showed that it was
possible to construct models in which the technifermions had only
isospin symmetric charges under the new $U(1)$.  They referred to such
models as ``natural" ${\rm TC}^2$ models.  In order to produce mixing
among heavy and light generations, such models seem to require direct couplings
of the new $U(1)$ to fermions in the first two generations \cite{lane}, and
thus predict a variety of effects which are potentially observable in
precision electroweak measurements\footnote{Since both top quarks and
technifermions
condense, these models also contain ``top-pions" \cite{kominis} which, if
light, can
produce observable effects.  Here we ignore ``top-pion" corrections;
since ``top-pion" loops further reduce $R_b$, including them could only worsen
the fit to data.}.

In this paper we perform global fits to precision data for three
examples of natural ${\rm TC}^2$ scenarios: a simple ``baseline"
scenario with universal couplings to quarks and leptons (of a sort in
which it does not seem too difficult to allow for mixing), an ``optimal"
scenario where it is assumed that there is no direct coupling to the
first two generations,
and an explicit model due to Lane \cite{lane} where new $U(1)$ charges
have been assigned to all the known fermions in such a way as to allow
for intergenerational mixing and gauge anomaly cancelation.  For the
``baseline'' and ``optimal'' scenarios, we find that the $Z^\prime$ must
typically weigh more than 2 TeV, although for the ``optimal" scenario the
$Z^\prime$ mass can be substantially smaller for a range of gauge
couplings. For Lane's model \cite{lane}, in which there are couplings
of ${\cal O}$(10), we find that the $Z^\prime$ must weigh more than
approximately 20 TeV.

\section{Electroweak Phenomenology of Top-Color Assisted Technicolor}
\label{sec:ewphen}
\setcounter{equation}{0}

The electroweak gauge symmetry in \tcii models is $SU(2)_L \times U(1)_1
\times U(1)_2$. Here $U(1)_2$ is the, presumably strong,
interaction with isospin-violating quark couplings (that allows for
top-quark, but not for bottom-quark, condensation) and $U(1)_1$ is a weak
gauge interaction.  The pattern of electroweak gauge symmetry
breaking that is required is more complicated than that in ordinary
technicolor models; it generally involves two scales (rather than just
one) to break the $SU(2)_L \times U(1)_1 \times U(1)_2$ symmetry down to
$U(1)_{em}$.  The required pattern of breaking is:
\begin{center}
$SU(2)_{L} \otimes U(1)_1 \otimes U(1)_2 $
\end{center}
\vspace{-20pt}
\begin{center}
$\downarrow \ \ \ \ \  u $
\end{center}
\vspace{-20pt}
\begin{center}
$SU(2)_{L} \otimes U(1)_Y$
\end{center}
\vspace{-20pt}
\begin{center}
$\downarrow\ \ \ \ \ v $
\end{center}
\vspace{-20pt}
\begin{center}
$U(1)_{em}$,
\end{center}
where hypercharge, $Y=Y_1 + Y_2$, is equal to the sum of the generators
of the two $U(1)$'s.

The gauge covariant derivative is given in the canonical basis by
\beq
\partial^\mu + ig \,T^a \,W_{a}^\mu + ig_1' \,Y_1\,B_1^\mu
+ig_2'\,Y_2 B_2^\mu ,
\eeq
where the $T^a$, $a=1$~to~3, are the generators of $SU(2)$.
The gauge couplings may be written
\beq
g ={e\over \st}\,,\quad
g_1' = {g^\prime\over\cp} = {e\over \cp\ct}\,,\quad
g_2' = {g^\prime\over\sp} = {e\over \sp\ct}\,,
\eeq
in terms of the usual weak mixing angle $\theta$ and a new mixing angle
$\phi$.
It is convenient to rewrite the neutral gauge bosons in terms of the photon,
\beq
A^\mu = \ct\,(\cp\, B^\mu_1 + \sp\, B^\mu_2) + \st\, W^\mu_3,
\eeq
which couples to electric charge $Q$ with strength $e$, a field
\beq
Z^\mu_1=-\st\,(\cp\, B^\mu_1 + \sp\, B^\mu_2)+\ct\, W^\mu_3,
\eeq
which couples as the standard model $Z$ would couple,
to $T_3 - Q \sin^2\theta$  with strength
\beq
{e\over \st\ct}\, ,
\eeq
and the field
\beq
Z^\mu_2 = -\sp\, B^\mu_1 + \cp\, B^\mu_2,
\eeq
which couples to the current $Y^\prime = Y_2 - \sin^2\phi Y$ with
strength
\beq
{g^\prime\over \sp\cp} = {e\over \ct\sp\cp}\, .
\eeq
In this basis, using the relation
\beq
Q=T_3 +Y
\label{Q}
\eeq
and the fact that $Q$ is conserved, the
mass-squared matrix for the $Z_1$ and $Z_2$ can be written as:
\beq
M_Z^2=\left({e  \over {2 \st \ct}} \right)^2\,
\pmatrix{<T_3 T_3>&
{ {\sin \theta }\over {\sp \cp}} <T_3 Y^\prime > \cr
{ {\sin \theta }\over {\sp \cp}} <T_3 Y^\prime >&
{{\sin^2 \theta}\over {\sin^2\phi \cos^2\phi}} <Y^\prime Y^\prime>\cr}\,,
\eeq
where, from the charged-$W$ masses we see that
$<T_3 T_3> = v^2 \approx (250\, {\rm GeV})^2$.

In natural \tcii models (where the strong $U(1)$ couplings to technifermions
are isospin symmetric) the expectation value leading to $Z_1 - Z_2$
mixing, \nobreak{$<T_3 Y^\prime>$}, can be calculated
{\it entirely} in terms of the gauge couplings,
$v$, and the $Y_2$ charges of the left- and right-handed top quark.
Using the definition of $Y^\prime$, we see that
\beq
<T_3 Y^\prime> \, = \, <T_3 Y_2> - \sin^2\phi <T_3 Y>\, .
\eeq
Since $Y=Q-T_3$ and $Q$ is conserved, the last term is equal to
$+\sin^2\phi <T_3 T_3>$. Furthermore in natural
\tcii models, since the technifermion $Y_2$-charges are assumed to be isospin
symmetric, the technifermions do not contribute to the first term. The
only contribution to the first term comes from the top-quark
condensate
\beq
{<T_3 Y_2> \over <T_3 T_3>} = 2(Y^{t_L}_2-Y^{t_R}_2){f^2_t \over v^2}\, ,
\eeq
where $f_t$ is the analog of $f_\pi$ for the top-condensate and
is equal to
\beq
f_t^2 \approx   {{N_c }\over{8\pi^2}}\, m_t^2
\log\left({{M^2}\over{m_t^2}}\right)
\eeq
in the Nambu---Jona-Lasinio \cite{NJL} approximation,
and $M$ is the top-gluon mass. For $m_t \approx 175$ GeV and $M\approx
1$ TeV, we find $f_t \approx 64$ GeV.

If we define
\beq
x \equiv {{\sin^2 \theta}\over {\sin^2\phi \cos^2\phi}}  \,
{{<Y^\prime Y^\prime>}\over{ <T_3 T_3>}}  \, \propto {u^2\over v^2} \, ,
\eeq
and
\beq
\epsilon \equiv   2  \,  {{f_t^2}\over{v^2}}
\left( Y_2^{t_L} - Y_2^{t_R} \right)\, ,
\eeq
the $Z_1 - Z_2$ mass matrix can be written as
\beq
M_Z^2=M^2_{Z}|_{\rm SM}\,
\pmatrix{
1 & {\tan\phi\sin \theta }\left(1 +{\epsilon\over\sin^2\phi} \right) \cr
{\tan\phi \sin \theta}  \left( 1 +{\epsilon\over\sin^2\phi}\right)  & x \cr}
\, .
\eeq
In the large-$x$ limit the mass eigenstates are
\beqa
Z  & \approx & Z_1 - {{\tan\phi \sin \theta }\over {x}} \left( 1 +
{{\epsilon}\over{\sin^2\phi}} \right)Z_2 \\
Z^\prime & \approx & {{\tan\phi \sin\theta }\over {x}} \left( 1 +
{{\epsilon}\over{\sin^2\phi}} \right) Z_1 + Z_2
\eeqa
The shifts in the $Z$ coupling to $f {\overline f}$ (with $e/(\cos\theta
\sin\theta)$ factored
out) are
therefore given by:
\beq
\delta g^f \approx -  \, {{\sin^2 \theta}\over{x \cos^2\phi}}  \, \left( 1 +
{{\epsilon}\over{\sin^2\phi}}\right)
\left[Y_2^f -\sin^2\phi Y^f \right]\, .
\label{dg}
\eeq
Mixing also shifts the $Z$ mass, and gives a contribution to
$T$ equal to:
\beq
\alpha T \approx { {\tan^2\phi \,\sin^2 \theta}\over {x}} \left(1 +
{{\epsilon}\over{\sin^2\phi}}\right)^2\, .
\label{T}
\eeq

The shifts in the $Z$-couplings and mass are sufficient to describe
electroweak phenomenology on the $Z$-peak. For low-energy processes,
in addition to these effects we must also consider the effects of
$Z^\prime$-exchange. To leading order in $1/x$, these effects may
be summarized by the four-fermion interaction
\beq
-{\cal L}^{Z^\prime}_{\rm NC} = {4 G_F \over \sqrt{2}}  \,
{\sin^2\theta \over x \sin^2\phi \cos^2\phi}  \,
\left(J_{Y_2} - \sin^2\phi J_Y \right)^2\, ,
\eeq
where $J_{Y_2}$ and $J_Y$ are the $Y_2$- and hypercharge-currents,
respectively.  It is useful to note that if $\epsilon$ is negative, then
all the $Z$ pole mixing effects (equations (\ref{dg}) and (\ref{T}))
vanish when $\sin^2\phi = -\epsilon$, although the low-energy effects of
$Z^\prime$ exchange do not.

The result of all these corrections is that the predicted values of
many electroweak observables are altered from those given by the
standard model\footnote{We are using $\alpha_{em}(M_Z)$, $G_F$, and
$M_Z$ as the tree-level inputs.} \cite{fit,unun,NClimits}. The shifts in the
predictions depend on the charge assignments of the quarks and leptons under
the new $U(1)$.
The ``baseline" scenario we consider in this paper has universal couplings of
the strong $U(1)$
to  quarks and leptons given by
$Y_2=Y$
(and $Y_1=0$ for the ordinary fermions).
This results in a shift in the total $Z$ width given by:
\begin{equation}
 \Gamma_Z = \left( \Gamma_Z \right)_{SM} \left(1 + 0.250  {{1} \over {x
\cos^2\phi }}
-0.017  {{1}\over{x \cos^2\phi \sin^2 \phi}} + 0.039  {{\tan^2\phi}\over{x}}
\right) ~.
\label{GZUO}
\end{equation}
The full list of changes to the electroweak observables for this
``baseline" scenario appears in the Appendix.

For the ``optimal" scenario we take the generation-dependent
charge assignments of $Y_2$ to have the
values of ordinary hypercharge on the third generation, and zero on the first
two generations.  This results, for example, in a shift in the total $Z$ width:
\begin{equation}
 \Gamma_Z = \left( \Gamma_Z \right)_{SM} \left(1 + 0.056  {{1} \over {x
\cos^2\phi}}
-4.0 \times 10^{-3} {{1}\over{x \cos^2\phi  \sin^2 \phi}} + 0.039
{{\tan^2\phi}\over{x}} \right)~.
\end{equation}

For Lane's model we use the charge assignments given in
the Appendix of ref. \cite{lane}. These charges are listed in Table
\ref{laneQ},
with a normalization implied by equation (\ref{Q}). The shift in the total $Z$
width in
Lane's model is given by:
\begin{equation}
\Gamma_Z = \left( \Gamma_Z \right)_{SM} \left(1 + 9.4 \times 10^{-3}  {{1}
\over {x \cos^2\phi}} -8.4 \times 10^{-3}  {{1}\over{x \cos^2\phi \sin^2 \phi}}
+ 0.039  {{\tan^2\phi}\over{x}} \right) ~.
\end{equation}

\begin{table}[htbp]
\begin{center}
\begin{tabular}{|l|l||l|l|}\hline\hline
1st, 2nd &  $Y_2$  & 3rd & $Y_2$ \\ \hline\hline
$(u,d)_L$,  $(c,s)_L$  & $-10.5833$ & $(t,b)_L$& $8.7666$  \\ \hline
$u_R$,  $c_R$  & $-5.78333$ & $t_R$& $11.4166$  \\ \hline
$d_R$,  $s_R$  & $-6.78333$ & $b_R$& $10.4166$  \\ \hline
$(\nu,e)_L$,  $(\nu,\mu)_L$  & $-1.54$ & $(\nu,\tau)_L$& $-1.54$  \\ \hline
$e_R$,  $\mu_R$  & $2.26$ & $\tau_R$& $2.26$  \\ \hline
\hline
\end{tabular}
\end{center}
\caption{The charges for Lane's model.  The first two columns refer to the
first two generations, while the last two columns refer to the third
generation.}
\label{laneQ}
\end{table}

\section{Comparison with Data}
\label{sec:data}
\setcounter{equation}{0}

\subsection{Data and the Standard Model}
\label{subsec:datasm}

Before describing the details of the fit, we briefly discuss higher-order
corrections. Beyond tree-level, the predictions of the
standard model (as well as predictions of extended models like those we
are discussing)  depend on the values of
$\alpha_s(M_Z)$ and the top-quark mass $m_t$. Given the success of the
standard model, we expect that, for the allowed range of  $1/x$,
the changes in the predicted values of
physical observables due to radiative corrections in the standard
model or extended models will be approximately the same
for the same values of $\alpha_s(M_Z)$ and $m_t$.

The best-fit standard model predictions \cite{langacker} which we use
are based on a top quark mass of 173 GeV (taken from a fit to
precision electroweak data) which is consistent with
the mass range preferred by
CDF  ($176\pm13$ GeV) and  D0 ($170 \pm 25$ GeV)\cite{cdf}.

The treatment of $\alpha_s(M_Z)$ is more problematic: the LEP
determination for $\alpha_s(M_Z)$ comes from a {\it fit} to
electroweak observables {\it assuming} the validity of the standard
model. For this reason it is important \cite{alphaZbb,unun,NClimits} to
understand
how the bounds vary for different values of $\alpha_s(M_Z)$. We
present  bounds  both
for $\alpha_s(M_Z) = 0.124$ (which is the LEP best-fit value assuming
the standard model is correct \cite{langacker}) and for
$\alpha_s(M_Z)=0.115$ (which is the most precisely measured
value, determined from  lattice results
\cite{lattice} and consistent with deep-inelastic scattering
\cite{langacker,deep}).
To the accuracy to which we work, the $\alpha_s$ dependence of the
standard model predictions only appears in the $Z$ partial widths and
we use \cite{langacker}
\beq
\Gamma_q = \Gamma_q|_{\alpha_s=0}\left(1+ {{\alpha_s}\over{\pi}}
+ 1.409 \left({{\alpha_s}\over{\pi}}\right)^2 -
12.77\left({{\alpha_s}\over{\pi}}\right)^3 \right)
\eeq
to obtain the standard model predictions for $\alpha(M_Z)=0.115$.

With the above caveats, we have performed a global fit for the
``baseline"  ${\rm TC}^2$ scenario, the
``optimal" ${\rm TC}^2$ scenario, and for Lane's model to all
precision
electroweak data: the $Z$ line shape, forward backward asymmetries,
$\tau$ polarization, and left-right asymmetry measured at LEP and SLC;
the $W$ mass measured at FNAL and UA2; the electron and neutrino
neutral current couplings determined by deep-inelastic scattering; the
degree of atomic parity violation measured in Cesium; and the ratio of
the decay widths of $\tau \to \mu \nu \bar\nu$ and $\mu\to e \nu \bar
\nu$.

\subsection{Results of  Fitting}
\label{subsec:fit}

Using the current experimental values\footnote{The experimental data and
standard model
predictions are given in Table \ref{Input} in the Appendix.} we have fit
the ``baseline" and ``optimal" ${\rm TC}^2$ scenarios and Lane's ${\rm TC}^2$
model
to the data.
Figure 1 summarizes the fit to the ``baseline" scenario (with universal
couplings
to quarks and leptons) by displaying the
95\%  confidence level\footnote{This corresponds to $\Delta \chi^2=4.0$
relative
to the best-fit point,
since we are fitting to one parameter $1/x$, holding $\sin^2\phi$ fixed.}
lower bound on the  $Z^\prime$ mass for different
values of $\sin^2 \phi$
($\alpha_s(M_Z) = 0.115$ corresponds to the
solid line and  $\alpha_s(M_Z) = 0.124$ to the dashed line). The plot was
created as follows: for each value of $\sin^2 \phi$ we fit to  $1/x$;
we then found the upper bound on $1/x$ and
translated it into a lower bound on the $Z^\prime$ mass.
The resulting bounds are quite
sensitive to $\alpha_s(M_Z)$, with $\alpha_s(M_Z)=0.124$ giving a much tighter
constraint on the mass for most values of $\sin^2\phi$.
The plot clearly shows a dip in the bound at
$\sin^2\phi = -\epsilon \approx 0.07$, where the $Z - Z^\prime$ mixing
vanishes.
For generic values of $\sin^2\phi$ the 95 \% bound is roughly 2
TeV. While the results shown are for  the case where $Y_2=Y$, we
expect that {\it any} model where the $Y_2$ couplings to the first and
second generations are ${\cal O}$(1) will give similar results.

\begin{figure}[htbp]
\centering
\epsfysize=3in
\hspace*{0in}
\epsffile{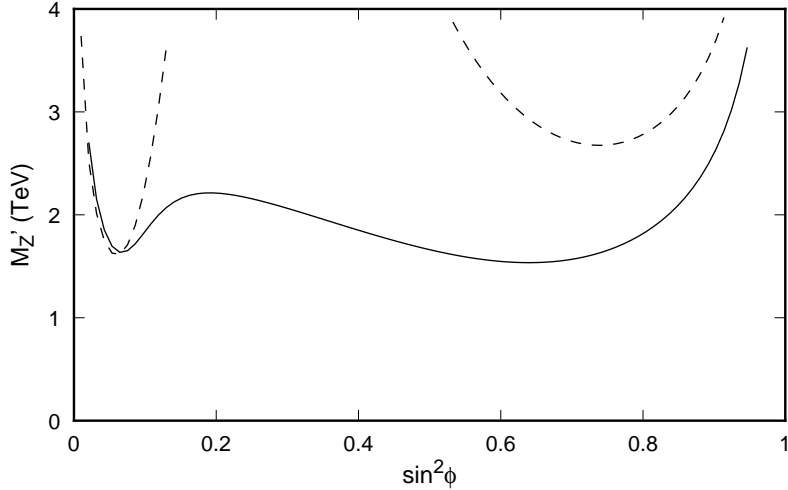}
\caption{The solid line is the 95\%
confidence lower bound for the mass of the $Z^\prime$
as a function of $\sin^2 \phi$  for the  ``baseline"  ${\rm TC}^2$ scenario,
using $\alpha_s(M_Z)=0.115$.   The dashed line
is for $\alpha_s(M_Z)=0.124$. }
\label{Fig1}
\end{figure}

Figure 2 summarizes the fit to the ``optimal" model with generation
dependent couplings by displaying the 95\% confidence level lower bound
(solid line) on the $Z^\prime$ mass for different values of $\sin^2
\phi$ (using $\alpha_s(M_Z)=0.115$; for this model the bounds are not
very sensitive to the value of $\alpha_s(M_Z)$).  Also shown is the 68\%
confidence level lower bound (dashed line) which displays the
sensitivity of the fit for different values of $\sin^2 \phi$.

\begin{figure}[htbp]
\centering
\epsfysize=3in
\hspace*{0in}
\epsffile{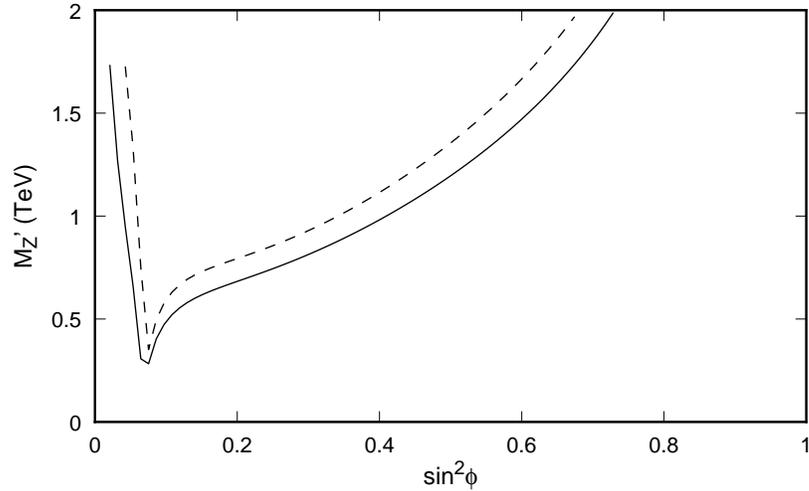}
\caption{The solid line is the 95\%
confidence lower bound for the mass of the $Z^\prime$
as a function of $\sin^2 \phi$  for the ``optimal" scenario
(using $\alpha_s(M_Z)=0.115$). The dashed line
is the 68\% confidence lower bound.}
\label{Fig2}
\end{figure}

For Lane's model the fit results are summarized in Figure 3, which displays the
95\% and 68\% confidence lower bounds on the $Z^\prime$ mass
(solid and dashed lines) as a function of $\sin^2 \phi$.
The lowest possible  $Z^\prime$ mass at the
95\% confidence level  is roughly 20 TeV.  The bound is significantly larger
in this model since the new $U(1)$ charges (as listed in Table \ref{laneQ})
are ${\cal O}(10)$ rather than ${\cal O}(1)$.  As anticipated by Lane
\cite{lane},
the fit for this model  is very sensitive to the
atomic parity violation measurement in Cesium, as can be seen from the
predicted value of the $Q_W(Cs) $ in this model:
\begin{equation}
 Q_W(Cs) = \left( Q_W(Cs) \right)_{SM} -324  {{1} \over {x \cos^2\phi}} -24887
{{1}\over{x \cos^2\phi \sin^2\phi}} + 16.7  {{\tan^2\phi}\over{x}} ~.
\end{equation}
The large coefficients result from the large quark charges in this model, and
the
large number of quarks in a Cesium atom. In a different version of Lane's model
\cite{laneunpub}, where the new $U(1)$ has vectorial couplings to leptons, the
95\% confidence level bound drops to 2.7 TeV, in accord with the results
of the ``baseline" scenario fit.

\begin{figure}[htbp]
\centering
\epsfysize=3in
\hspace*{0in}
\epsffile{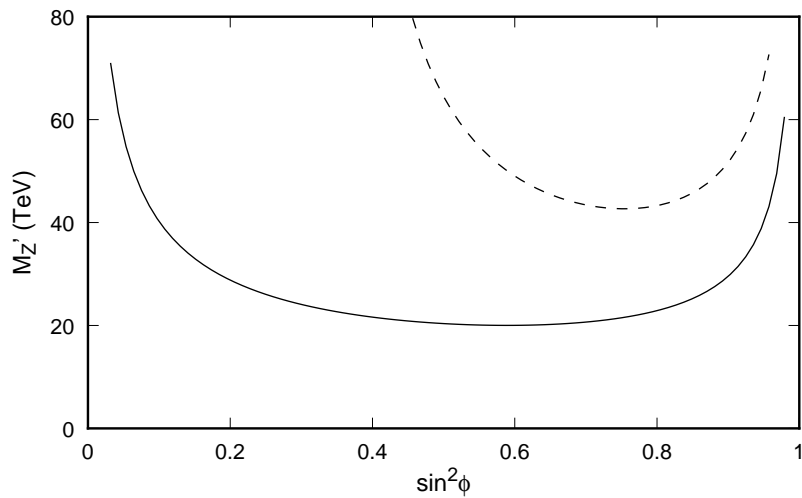}
\caption{The solid line is the 95\%
confidence lower bound for the mass of the $Z^\prime$
as a function of $\sin^2 \phi$  for Lane's ${\rm TC}^2$ model
(using $\alpha_s(M_Z)=0.115$).   The dashed line
is the 68\% confidence lower bound.}
\label{Fig3}
\end{figure}

The summary of the quality of the fits is given in Table
\ref{Fit115} for two different values of $\sin^2 \phi$ for both the
``baseline" and ``optimal" scenarios.  We first show the result of allowing
both
$\sin^2\phi$ and $1/x$ to vary.  For the ``baseline" model the fit prefers
$\sin^2\phi$
to be close to 1, however this makes the coupling $g_1^\prime$ diverge.
Requiring
$g_1^\prime /4 \pi  \le 1$ gives a best fit at $\sin^2 \phi=0.99$, which
corresponds to
a $Z^\prime$ mass of  $M_{Z^\prime} =11.6 ^{+ 3.4}_{-1.8}$ TeV.  For the
``optimal" model the fit yields $\sin^2\phi=0.62$ which corresponds to
$M_{Z^\prime} =2.1^{+0.6}_{-0.3}$ TeV.

Such large values of $\sin^2\phi$ imply that the new $U(1)$ gauge coupling is
quite weak, and  therefore such models require a fine-tuning to arrange for a
top condensate
but not a bottom condensate \cite{tc2}.  Thus we show a second fit where
we required that $g_2^\prime/4 \pi >0.1$ (which corresponds to $\sin^2\phi <
0.1$).
Given this constraint we chose $\sin^2\phi$ such that the best-fit value for
$1/x$
corresponded to the lightest  possible mass for the $Z^\prime$.
For the  ``baseline" model this occurs at $\sin^2\phi=0.036$ which corresponds
to $M_{Z^\prime} = 4.4^{+\infty}_{-1.8}$ TeV,
while for the ``optimal" model the minimum is at $\sin^2\phi=0.07$, where
there is no $Z$ - $Z^\prime$ mixing, which
corresponds
to  $M_{Z^\prime} = 290^{+\infty}_{-110}$ GeV.
A fit for Lane's model \cite{lane} is not shown since there is no value
of $\sin^2 \phi$ that has a best fit with a physical  $Z^\prime$
mass  (i.e. $1/x > 0$).

Table \ref{Fit115} shows the fit to the standard model for
comparison.  The
percentage quoted in the Table is the probability of obtaining a
$\chi^2$ as large or larger than that obtained in the fit, for the
given number of degrees of freedom (df), assuming that the model is
correct.  Thus a small probability corresponds to a poor fit, and a
bad model.  Note that the standard model gives a poor fit to the data,
while the fine-tuned ${\rm TC^2}$ scenarios are only modestly better, and in
the absence of fine-tuning the  ${\rm TC^2}$ scenarios are slightly worse.

\begin{table}[htbp]
\begin{center}
\begin{tabular}{|l||c||c|c|c||c|}\hline\hline
Model &  $\sin^2\phi$ & $\chi^2$&df&$\chi^2/{\rm df}$  & probability \\
\hline\hline
standard model  & - & $43.7$ & $23$& $1.90$ & $0.6\%$ \\ \hline
``baseline"   & 0.99 &$37.5$ & $21$ & $1.79$ & $1.5\%$ \\ \hline
``optimal"    & 0.62 &$37.3$ & $21$ & $1.78$ & $1.6\%$ \\ \hline
``baseline", no tuning  & 0.036 &$43.4$ & $22$ & $1.97$ & $0.4\%$ \\ \hline
``optimal", no tuning  & 0.069 &$43.3$ & $22$ & $1.97$ & $0.4\%$ \\ \hline
\hline
\end{tabular}
\end{center}
\caption{The best fits for the standard model, the  ``baseline" ${\rm TC^2}$
scenario, and the ``optimal" ${\rm TC^2}$ scenario, for different values of
$\sin^2 \phi $, assuming
$\alpha_s(M_Z)=0.115$. $\chi^2$ is the sum
of the squares of the difference between prediction and experiment,
divided by the error, the number of degrees of freedom, df, is the number of
experiments
minus the number of fitted parameters. The phrase "no tuning" corresponds to
the
requirement $\sin^2\phi < 0.1$ as discussed in the text.}
\label{Fit115}
\end{table}

For comparison we have also performed the fits using
$\alpha_s(M_Z)=0.124$; the quality of the fit is summarized in Table
\ref{Fit124}.  The quality of all the fits improves, but there are only
small changes in the relative goodness of fit.  The best fit for the
``baseline" model corresponds to an unphysical
$Z^\prime$ mass (i.e. $1/x < 0$). If we require a physical  $Z^\prime$ mass
then the fit prefers  $\sin^2\phi$ close to zero, so no fine-tuning is
required. For $\sin^2\phi = 0.01$
we find $M_{Z^\prime}=5.9 ^{+5.2}_{-1.4}$ TeV.
The best fit for the ``optimal" model corresponds to $\sin^2\phi = 0.99$
and $M_{Z^\prime}=18.9 ^{+8.6}_{-3.6}$ TeV.
We also show a fit  for the ``optimal" model where we require the absence of
fine-tuning
($\sin^2\phi <0.1$), and the lightest possible $Z^\prime$ mass.  We found this
occurs for $\sin^2\phi=0.07$, the value where there is no $Z$ - $Z^\prime$
mixing,
which corresponds
to  $M_{Z^\prime}=1.1^{+\infty}_{-0.75}$ TeV.
Again we do not show a fit for Lane's model  since there is no value
of $\sin^2 \phi$ that has a best fit with a physical  $Z^\prime$ mass.
Here the ``optimal" scenario provides a  better fit than the standard model if
fine-tuning
is allowed.  In the absence of fine-tuning the ${\rm TC^2}$ scenarios give
fits to the data that are slightly worse than the standard model.

\begin{table}
\begin{center}
\begin{tabular}{|l||c||c|c|c||c|}\hline\hline
Model &  $\sin^2\phi$ & $\chi^2$&df&$\chi^2/{\rm df}$  & probability \\
\hline\hline
standard model & - & $33.9$ & $22$& $1.47$ & $6.7\%$ \\ \hline
``baseline", no tuning  & 0.01 &$31.9$ & $21$ & $1.52$ & $5.9\%$ \\ \hline
``optimal"  & 0.99  &$30.3$ & $21$ & $1.38$ & $11.2\%$ \\ \hline
``optimal", no tuning  & 0.071 &$33.9$ & $22$ & $1.54$ & $5.1\%$ \\ \hline
\hline
\end{tabular}
\end{center}
\caption{The best fits for the standard model, the  ``baseline" ${\rm TC^2}$
scenario, and the ``optimal" ${\rm TC^2}$ scenario, for different values of
$\sin^2 \phi $, assuming
$\alpha_s(M_Z)=0.124$. For this value of $\alpha_s(M_Z)$ the best fit for the
``baseline" scenario with a physical $Z^\prime$ mass requires no fine-tuning.
See Table 2 for an explanation of the notation.}
\label{Fit124}
\end{table}

\section{Conclusions}
\label{sec:concl}

In this paper we have used precision electroweak measurements to constrain
``natural" ${\rm TC}^2$ models.  The models we examined can provide a better
fit to
data than the standard model if fine-tuning is allowed, however in the absence
of
fine-tuning they generally provide a fit that is slightly worse than the poor
fit given by
the standard model.
Generically we have found that the mass of the new
$Z^\prime$ in such models has to be above roughly 2 TeV.   In models with no
couplings of the new $U(1)$ to the first two generations, the bound on
the $Z^\prime$ mass can be lower, but it seems difficult
to allow for mixing between generations in such models. In the explicit model
due to
Lane \cite{lane}, which allows for inter-generational mixing, the bounds
are much more stringent, the couplings to the first two generations are
${\cal O}$(10), requiring a $Z^\prime$ mass above 20 TeV.


\vspace{12pt} \centerline{\bf Acknowledgments} \vspace{12pt}

We thank K. Lane for six months of constant nudging and for comments on the
manuscript.  R.S.C. acknowledges the support of an NSF
Presidential Young Investigator Award.  J.T.  acknowledges the support
of a Japan Society for the Promotion of Science Fellowship. {\em This
work was supported in part by the National Science Foundation under
grants PHY-9057173 and PHY-9501249, and by the Department of Energy
under grant DE-FG02-91ER40676.}

\appendix
\section{Appendix: Data and the ``baseline" ${\rm TC}^2$ scenario predictions}
\label{sec:basselineapp}
\setcounter{equation}{0}
\begin{table}[htbp]
\begin{center}
\begin{tabular}{|c|l|l|l|}\hline\hline
Quantity & Experiment & SM \\
\hline \hline
$\Gamma_Z$ & 2.4964 $\pm$ 0.0022 & 2.4925   \\
$R_e$ & 20.801 $\pm$ 0.058 & 20.717   \\
$R_\mu$ & 20.795 $\pm$ 0.043 & 20.717   \\
$R_\tau$ & 20.815 $\pm$ 0.061 & 20.717   \\
$\sigma_h$ & 41.490 $\pm$ 0.078 & 41.492   \\
$R_b$ & 0.2215 $\pm$ 0.0017 & 0.2156   \\
$R_c$ & 0.1596 $\pm$ 0.0070 & 0.1720   \\
$A_{FB}^e$ & 0.0152 $\pm$ 0.0027 & 0.0155   \\
$A_{FB}^\mu$ & 0.0163 $\pm$ 0.0015 & 0.0155   \\
$A_{FB}^\tau$ & 0.0203 $\pm$ 0.0022 & 0.0155   \\
$A_{\tau}(P_\tau)$ & 0.1394 $\pm$ 0.0069 & 0.1440   \\
$A_{e}(P_\tau)$ & 0.1429 $\pm$ 0.0079 & 0.1440   \\
$A_{FB}^b$ & 0.1002 $\pm$ 0.0028 & 0.1010   \\
$A_{FB}^c$ & 0.0756 $\pm$ 0.0051 & 0.0720   \\
$A_{LR}$ & 0.1551 $\pm$ 0.0040 & 0.1440   \\
$M_W$ & 80.17 $\pm$ 0.18 & 80.34   \\
$M_W/M_Z$ & 0.8813 $\pm$ 0.0041 & 0.8810   \\
$g_L^2(\nu N \rightarrow \nu X)$ & 0.3003 $\pm$ 0.0039 & 0.3030   \\
$g_R^2(\nu N \rightarrow \nu X)$ & 0.0323 $\pm$ 0.0033 & 0.0300   \\
$g_{eA}(\nu e \rightarrow \nu e)$ & -0.503 $\pm$ 0.018 & -0.507   \\
$g_{eV}(\nu e \rightarrow \nu e)$ & -0.025 $\pm$ 0.019 & -0.037   \\
$Q_W(Cs)$ & -71.04 $\pm$ 1.81 & -72.88   \\
$R_{\mu \tau}$ & 0.9970 $\pm$ 0.0073 & 1.0   \\
\hline\hline
\end{tabular}
\end{center}
\caption{Experimental \protect\cite{langacker,LEP,taudec} and
predicted values of
electroweak observables for the standard model  for $\alpha_s(M_Z)=0.115$.
The standard model values
correspond to the best-fit values (with $m_t=173$ GeV, $m_{\rm Higgs}
= 300$ GeV) in \protect\cite{langacker}, with
$\alpha_{em}(M_Z)=1/128.9$,  and corrected for the change in
$\alpha_s(M_Z)$.}
\label{Input}
\end{table}

The following equations are the predictions of the ``baseline" scenario as
functions of $\phi$ and $x$. For brevity we use the notation $c = \cos\phi$,
$s=\sin\phi$, and $t=\tan\phi $.

\begin{equation}
 \Gamma_Z = \left( \Gamma_Z \right)_{SM} \left(1 + 0.250  {{1} \over {x c^2}}
-0.017  {{1}\over{x c^2 s^2}} + 0.039  {{t^2}\over{x}} \right)
\end{equation}

\begin{equation}
 R_e = \left( R_e \right)_{SM} \left(1 -0.160  {{1} \over {x c^2}} + 9.9 \times
10^{-3} {{1}\over{x c^2 s^2}} + 0.201  {{t^2}\over{x}} \right)
\end{equation}

\begin{equation}
 R_\mu = \left( R_\mu \right)_{SM} \left(1 -0.160  {{1} \over {x c^2}} + 9.9
\times 10^{-3} {{1}\over{x c^2 s^2}} + 0.201  {{t^2}\over{x}} \right)
\end{equation}

\begin{equation}
 R_\tau = \left( R_\tau \right)_{SM} \left(1 -0.160  {{1} \over {x c^2}} + 9.9
\times 10^{-3} {{1}\over{x c^2 s^2}} + 0.201  {{t^2}\over{x}} \right)
\end{equation}

\begin{equation}
 \sigma_h = \left( \sigma_h \right)_{SM} \left(1 + 0.012  {{1} \over {x c^2}}
-7.3 \times 10^{-4} {{1}\over{x c^2 s^2}} -0.015  {{t^2}\over{x}} \right)
\end{equation}

\begin{equation}
 R_b = \left( R_b \right)_{SM} \left(1 + 0.036  {{1} \over {x c^2}} -2.2 \times
10^{-3} {{1}\over{x c^2 s^2}} -0.044  {{t^2}\over{x}} \right)
\end{equation}

\begin{equation}
 R_c = \left( R_c \right)_{SM} \left(1 -0.073  {{1} \over {x c^2}} + 4.5 \times
10^{-3} {{1}\over{x c^2 s^2}} + 0.094  {{t^2}\over{x}} \right)
\end{equation}

\begin{equation}
 A_{FB}^e = \left( A_{FB}^e \right)_{SM} -0.375  {{1} \over {x c^2}} + 0.023
{{1}\over{x c^2 s^2}} + 0.477  {{t^2}\over{x}}
\end{equation}

\begin{equation}
 A_{FB}^\mu = \left( A_{FB}^\mu \right)_{SM} -0.375  {{1} \over {x c^2}} +
0.023  {{1}\over{x c^2 s^2}} + 0.477  {{t^2}\over{x}}
\end{equation}

\begin{equation}
 A_{FB}^\tau = \left( A_{FB}^\tau \right)_{SM} -0.375  {{1} \over {x c^2}} +
0.023  {{1}\over{x c^2 s^2}} + 0.477  {{t^2}\over{x}}
\end{equation}

\begin{equation}
 A_{\tau}(P_\tau) = \left( A_{\tau}(P_\tau) \right)_{SM} -1.57  {{1} \over {x
c^2}} + 0.097  {{1}\over{x c^2 s^2}} + 2.00  {{t^2}\over{x}}
\end{equation}

\begin{equation}
 A_{e}(P_\tau) = \left( A_{e}(P_\tau) \right)_{SM} -1.57  {{1} \over {x c^2}} +
0.097  {{1}\over{x c^2 s^2}} + 2.00  {{t^2}\over{x}}
\end{equation}

\begin{equation}
 A_{FB}^b = \left( A_{FB}^b \right)_{SM} -1.12  {{1} \over {x c^2}} + 0.069
{{1}\over{x c^2 s^2}} + 1.42  {{t^2}\over{x}}
\end{equation}

\begin{equation}
 A_{FB}^c = \left( A_{FB}^c \right)_{SM} -0.877  {{1} \over {x c^2}} + 0.054
{{1}\over{x c^2 s^2}} + 1.11  {{t^2}\over{x}}
\end{equation}

\begin{equation}
 A_{LR} = \left( A_{LR} \right)_{SM} -1.57  {{1} \over {x c^2}} + 0.097
{{1}\over{x c^2 s^2}} + 2.00  {{t^2}\over{x}}
\end{equation}

\begin{equation}
 M_W = \left( M_W \right)_{SM} \left(1 -0.022  {{1} \over {x c^2}} + 7.6 \times
10^{-4} {{1}\over{x c^2 s^2}} + 0.166  {{t^2}\over{x}} \right)
\end{equation}

\begin{equation}
 M_W/M_Z = \left( M_W/M_Z \right)_{SM} \left(1 -0.022  {{1} \over {x c^2}} +
7.6 \times 10^{-4} {{1}\over{x c^2 s^2}} + 0.166  {{t^2}\over{x}} \right)
\end{equation}

\begin{equation}
 g_L^2(\nu N \rightarrow \nu X) = \left( g_L^2(\nu N \rightarrow \nu X)
\right)_{SM} + 0.117  {{1} \over {x c^2}} -2.6 \times 10^{-3} {{1}\over{x c^2
s^2}} + 0.058  {{t^2}\over{x}}
\end{equation}

\begin{equation}
 g_R^2(\nu N \rightarrow \nu X) = \left( g_R^2(\nu N \rightarrow \nu X)
\right)_{SM} -0.040  {{1} \over {x c^2}} + 0.055  {{1}\over{x c^2 s^2}} -0.020
{{t^2}\over{x}}
\end{equation}

\begin{equation}
 g_{eA}(\nu e \rightarrow \nu e) = \left( g_{eA}(\nu e \rightarrow \nu e)
\right)_{SM} -1.2 \times 10^{-3} {{1} \over {x c^2}} -0.101  {{1}\over{x c^2
s^2}} -7.3 \times 10^{-4} {{t^2}\over{x}}
\end{equation}

\begin{equation}
 g_{eV}(\nu e \rightarrow \nu e) = \left( g_{eV}(\nu e \rightarrow \nu e)
\right)_{SM} -0.311  {{1} \over {x c^2}} + 0.324  {{1}\over{x c^2 s^2}} -0.153
{{t^2}\over{x}}
\end{equation}

\begin{equation}
 Q_W(Cs) = \left( Q_W(Cs) \right)_{SM} + 34.3  {{1} \over {x c^2}} -51.3
{{1}\over{x c^2 s^2}} + 16.7  {{t^2}\over{x}}
\end{equation}

\begin{equation}
 R_{\mu \tau} = \left( R_{\mu \tau} \right)_{SM}
\end{equation}







\end{document}